\documentclass[twocolumn]{revtex4}
\usepackage{epsf}
\usepackage{graphicx}           

\begin{document}          

\title{
Quantum information analysis of electronic states at different molecular structures
}

\author{G. Barcza and \"O.~Legeza} 
\affiliation{Fachbereich Physik,
  Philipps-Universit\"at Marburg, 35032 Marburg, Germany\\
  Research Institute for Solid State Physics, H-1525 Budapest, P.\ O.\ Box 49, Hungary}

\author{K.\ H.\ Marti and M.\ Reiher} 
\affiliation{
Laboratorium f\"ur Physikalische Chemie, ETH Zurich, 8093 Zurich, Switzerland
}

\date{
Wed Aug 25 10:58:17 CEST 2010
}

\vskip -8pt
\begin{abstract}
We have studied transition metal clusters from a quantum information
theory perspective using the density-matrix renormalization group (DMRG) method.
We demonstrate
the competition between entanglement and interaction localization. We also
discuss the
application of the configuration interaction based dynamically extended active space procedure 
which significantly reduces the effective system size and accelerates the speed
of convergence for complicated molecular electronic structures to a great extent.
Our results indicate the importance of taking entanglement among molecular
orbitals into account in order to devise an optimal orbital ordering and carry out efficient 
calculations on transition
metal clusters. We propose a recipe to perform DMRG calculations in a black-box
fashion and we point out the connections of our work to other tensor network
state approaches.
\end{abstract}


\maketitle

\section{Introduction}

The most important property of multi-component quantum systems is entanglement
which corresponds to quantum correlations between particles or a collection of particles 
forming a larger subsystem~\cite{luigi2008}. The degree of entanglement is decisive for the behavior of 
all multi-component quantum systems and for the numerical algorithms developed to 
simulate such systems~\cite{white92,white92b,verstraetecirac04,murgverstraete07,murgverstraete08,verstraeteciracmurg08,vidal06,changlani09,marti10,murg10}. 

The concepts of quantum information theory (QIT) have been introduced to the
density matrix renormalization group (DMRG) method~\cite{white92,white92b}
and created a
fresh impetus to the development of new methods that focus on entanglement
optimization~\cite{vidal03,legeza03}.
It was shown that the quantum information entropy is
a direct measure of the behavior of quantum systems. Thus, it is mandatory
to develop 
methods and techniques to obtain as much information as possible 
from 
entropy analyses~\cite{legeza03,legeza04,rissler06,legeza05a,legeza06,affleck2009}.

Since the first application of the DMRG algorithm
to quantum chemical systems using the full electronic Hamiltonian~\cite{white99}, the method has gone through major algorithmic developments
by various groups~\cite{schollwoeck04,chan07,marti2010a,marti2010b,moritz07}.
In 2002, two groups independently provided the first large scale calculations on
diatomic molecules~\cite{chan02,legeza02a}.
Calculations on the ${\rm F}_2$ molecule by keeping 
more than
2000 block states (i.e., states defined on the active subsystem in DMRG) denoted
as $M$ in the following, and on the water molecule using 6000 block states~\cite{chan03}
confirmed that the 
DMRG method is capable of reaching the limit of the full-CI method in cases
where most standard 
quantum chemical approaches fail and cannot be applied due to the requirement of very large active spaces.
Efficient calculations of excited states and the relation between the 
DMRG wave function and the standard CI-expansion 
have already been discussed in Ref.~[{\onlinecite{legeza02b}]. The authors have also shown
that the DMRG wave function is also suited to study problems when 
the characteristics of the
wave function changes drastically, for example, when the bond length between diatomic
molecules is changed.
Therefore, it became evident that very accurate results can be obtained by
increasing the computational resources related to the DMRG block states.
On the other hand, the
more important questions, namely what is the minimum computational effort to
obtain results with a given accuracy is still unsolved and part of active 
research~\cite{legeza04,legeza05a,luo10,marti2010a}.

A key-ingredient of the DMRG
method related to its multi-component subsystem wave function representation is
entanglement.
A controlled manipulation of it is expected to extend capabilities of DMRG to
treat larger systems in a more efficient way. The development of 
entanglement reduction algorithms (ERA) is thus very appealing. 
Matrix product states (MPS)~\cite{oestlund95} that are inherently produced by
the DMRG algorithm can be used to localize the entanglement
by reordering sites~\cite{legeza03,chan02,legeza02a,moritz05,rissler06} or by
optimization of the
basis~\cite{white02,legeza05a,yanai06,yanai10,zgid08,zgid08a,ghosh08,luo10,murg10}.
Considering the
entanglement between arbitrarily chosen pairs of
sites~\cite{rissler06,marti10},
a network topology can be determined for tensor-network-state (TNS) algorithms~\cite{murg10}.
Further gain in speed of convergence can be achieved by an optimization of the
initialization procedure~\cite{legeza03,moritz06} in which 
highly entangled states are taken into account from the very beginning.
One such algorithm  
is the dynamically extended active space (DEAS) method~\cite{legeza03}, and
its extension by including CI-expansion techniques (CI-DEAS)~\cite{legeza04b,legeza10a}.
In this article, we will show that the proposed entropy-based optimization
scheme including the (CI-)DEAS 
and the dynamic block state selection (DBSS) 
procedures are vital ingredient to
obtain highly accurate results with tremendous savings in computation resources and
time. In addition, it is a smart way of reducing the Hilbert space
which allows us to study large active spaces with a modest number of
renormalized states in contrast to inefficient and expensive brute-force
DMRG calculations at a fixed number of renormalized states.
DMRG calculations on complex chemical systems using quantum information
theory have not yet been carried out, and are the subject of this work.

The very challenging binuclear oxo-bridged copper clusters identified in
Ref.~\cite{cram06} as a very difficult case for complete-active-space-like
calculations pose an ideal test for the
DMRG algorithm. In a pioneering study, some of us
investigated the DMRG algorithm for the prediction of relative energies 
of transition metal clusters of different molecular structures, the
bis($\mu$-oxo) and $\mu-\eta^2:\eta^2$ peroxo isomers of
[Cu$_2$O$_2$]$^{2+}$~\cite{marti08}. Based on these results, 
we drew the conclusion that the DMRG algorithm is 
suited to study transition metal complexes and clusters. 
Kurashige and Yanai have picked up that same problem as well 
and performed 
massively parallel DMRG calculations employing 2400 renormalized basis states
but on a larger active space with a smaller one-particle basis~\cite{kura09}. 
In their study, a new aspect
was the inclusion of a perturbative correction introduced originally by
White~\cite{white2005} for the one-site DMRG algorithm to prevent convergence to local minima
during the optimization process.
In the present study of [Cu$_2$O$_2$]$^{2+}$, we show that 
no convergence acceleration technique
such as white noise~\cite{chan02} or perturbative correction~\cite{kura09} 
is needed if an entropy-based optimization scheme is employed.

The purpose of the paper is to show how quantum information entropies can be
efficiently exploited to perform accurate calculations on complicated chemical
systems such as transition metal clusters. 
Our results indicate that -- within chemical accuracy --
much larger molecular systems
can be studied than before using very little computational resources.
In Sec.~III, we briefly discuss the various technical aspects of our
calculations 
while the entropy analysis of the
two isomers are presented in Sec.~IV. In Sec.~V, extensions to tensor network state methods
are discussed and our conclusions are given.  

\section{The Challenge of Binuclear Copper Clusters}

The reliable first-principles description of transition metal complexes and
clusters remains an important task for theoretical molecular
physics and quantum chemistry --- especially because of the
role of such molecules in catalysis and bioinorganic
chemistry~\cite{reih09,reih07fd,reih10fd,marti2010c}.

In 2006, Cramer~\emph{et al.} investigated several theoretical models on the
[Cu$_2$O$_2$]$^{2+}$ torture track and found incisive
discrepancies between them~\cite{cram06}. The CASSCF calculations even yield a
qualitatively wrong interpretation of the energy difference between the two isomers. 
The reason for this striking failure
lies in the inability of the CASSCF method to include all relevant molecular orbitals into the
one-particle active space that would be necessary to obtain a qualitatively correct
description of the electronic structure. 
Already the qualitative picture of an extended 
H\"uckel calculation indicated the requirement of more than doubling the active space for
binuclear transition metal clusters compared to the mononuclear
analog~\cite{marti08}.
An additional support for the fact that a very large active space is needed for a correct
description of the electronic structure of transition metals is the finding of
Pierloot~\emph{et al.} to include a second $d$ shell 
to obtain accurate results~\cite{pier01}.
As we showed in our previous work~\cite{marti08}, the DMRG algorithm is an ideal
candidate to tackle the description of challenging electronic
structures that require large active spaces occurring in transition metal chemistry.

This work reinvestigates the dicopper clusters with an improved 
methodology transforming the DMRG approach eventually into a black-box method 
with significantly reduced computational requirements.
For our DMRG calculations, the same active space, one-particle basis set and effective core
potential as in Ref.~\cite{marti08} was used.
All quantities calculated in this paper will be given in Hartree atomic units of which the
energy unit is one Hartree.

\section{Numerical procedure}

In this section, we outline the procedures and methods needed for the
entropy-based optimization scheme to efficiently carry out DMRG calculations.
Examples and figures will be presented for the $\mu-\eta^2:\eta^2$ peroxo isomer of [Cu$_2$O$_2$]$^{2+}$
while for the bis($\mu$-oxo) isomer only the final results are given. 

\subsection{Molecular Hamiltonian in second quantization}

In quantum chemical (QC-)DMRG applications, the electron--electron repulsion
is taken into account by an iterative
procedure that minimizes the Rayleigh quotient corresponding
to the electronic Hamiltonian given by
\begin{equation} \label{eqn:Hquantchem}
H = \sum_{i j \sigma} T_{i j} c_{i \sigma}^{\dagger} c_{j \sigma} +
\sum_{i j k l \sigma \sigma'} V_{i j k l} c_{i \sigma}^{\dagger} c_{j \sigma'}^{\dagger} c_{k \sigma'} c_
{l \sigma}
\end{equation}
and thus determines the full-CI wave function.
In Eq.~(\ref{eqn:Hquantchem}),
$c_{i \sigma}$ and $c_{i \sigma}^{\dagger}$ are the usual electron annihilation and creation operators,
$T_{i j}$ denotes the one-electron integral
comprising the kinetic energy of the electrons and the external
electric field of the nuclei. $V_{i j k l}$ stands for the two-electron
integrals and contains the electron--electron repulsion operator,
defined as
\begin{equation}
V_{i j k l} = \int d^3 x_1 d^3 x_2
\frac{ 
\phi_i^{*} (\vec{x}_1) \phi_j^{*} (\vec{x}_2) \phi_k (\vec{x}_2) \phi_l (\vec{x}_1) 
}{
|\vec{x}_1-\vec{x}_2|
} \, .
\end{equation}
We obtain the Hartree--Fock orbitals in a given one-particle basis of atomic Gaussian
functions and transform the one-electron and two-electron integrals in the
atomic basis set
to the Hartree--Fock molecular orbital basis using
the {\sc Molpro} program package~\cite{molpro_brief}, which we also employ to obtain
reference complete-active-space self-consistent field (CASSCF) and
complete-active-space configuration-interaction (CASCI) energies. 

In the QC-DMRG algorithm, a one-dimensional
chain is built up from molecular orbitals obtained from a suitable
mean-field or multi-configuration self-consistent field (MCSCF) calculation.
As will be discussed later, the one-orbital entropy function~\cite{legeza03}
and the two-orbital mutual information~\cite{rissler06}
provide a good starting configuration.
The irreducible representations of
the orbitals can also be used in the
DMRG procedure to carry out calculations for a given point-group symmetry~\cite{legeza03}.
This will be used in the present work, and
as an example, the components $T_{ij}$ of the one-electron operator are 
shown in Fig.~\ref{fig:cu2peroxo_T} for the $\mu-\eta^2:\eta^2$ peroxo isomer of
[Cu$_2$O$_2$]$^{2+}$. 
\begin{figure}
\centerline{
\includegraphics[scale=0.68]{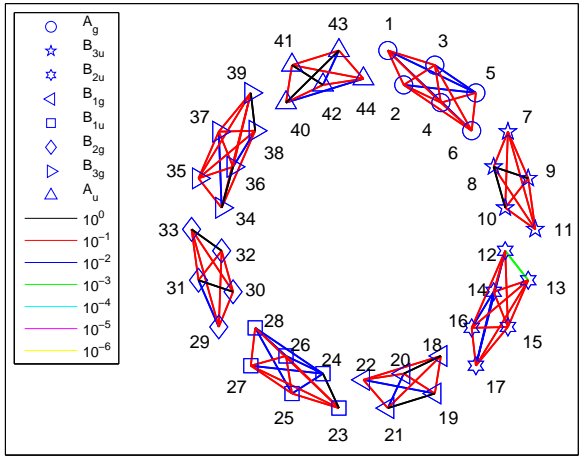}
}
\vskip -.5cm
\caption{(Color online) Pictorial representation of the magnitude of components
	$T_{ij}$ of the one-electron operator for the 
$\mu-\eta^2:\eta^2$ peroxo isomer of 
[Cu$_2$O$_2$]$^{2+}$. 
For better visibility the one-dimensional orbital chain is plotted on a 
circle with modulated radius in a clockwise direction. Orbitals belonging to different 
irreducible representations
are shown by different symbols. Numbers next to the symbols label molecular
orbitals.
\label{fig:cu2peroxo_T}
}
\end{figure}

\subsection{Two-site variant of the DMRG method}

For our DMRG calculations, we employ the two-site variant of the DMRG method introduced
by White~\cite{white92}. In the superblock configuration, two sites are 
between two blocks (sets) of orbitals.
To fix the notation
for the rest of the paper, we label the block states of
the left block containing $l$ orbitals, the two intermediate sites and the right block with
$r$ orbitals by $M_l, q_{l+1}, q_{l+2}, M_r$, respectively. The total number of orbitals $N$ is $l+2+r$.
When a larger block is formed during the renormalization steps by adding a
single site to the left or to the right  
block the new block states are denoted by $M_L$ and $M_R$ 
where $L=l+1$ and $R=r+1$.
The one-site variant of the DMRG~\cite{white2005} related to the MPS approach was first applied
in quantum chemistry in Ref.~[\onlinecite{legeza02a}] for fixed number
of block states but convergence properties were found considerably worse than
for the two-site DMRG algorithm.
This can be improved by the introduction of a local density operator~\cite{white2005}
which was used by Kurashige and Yanai for the calculation of the transition
metal cluster compound with
2400 block states~\cite{kura09}. Although the size of superblock Hilbert space is smaller
in this case, we use the two-site variant since this configuration allows one to
control the number
of block states dynamically and convergence to local minima is less likely~\cite{legeza02a}.

\subsection{Dynamic Block State Selection}

A fundamental concept related to the inseparability and
non-locality of quantum mechanics is entanglement.
Since the QC-DMRG algorithm approximates a composite system with long-range interactions, the
results of quantum information theory
can be used to understand the criteria of its convergence. 

The two-site variant of the DMRG method has originally been employed with a fixed
number of block states while the degree of entanglement between the DMRG blocks
for a given superblock configuration is related to the Schmidt number $r_{\rm Schmidt}$. 
For a pure target state 
$|\Psi_{\rm TS} \rangle \in {\Lambda} = {\Lambda}_L\otimes {\Lambda}_R$, with $\dim {\Lambda}_L=M_L$,
 $\dim {\Lambda}_R=M_R$,
the Schmidt decomposition states that 
\begin{equation}
|\Psi_{\rm TS} \rangle = \sum_{i=1}^{r_{\rm Schmidt}\leq\min(M_L,M_R)} \omega_i |e_i\rangle \otimes |f_i\rangle,
\label{eq:schmidt}
\end{equation}
where $|e_i \rangle \otimes |f_i \rangle $ form a bi-orthogonal basis
$\langle e_i | e_j \rangle = \langle f_i | f_j \rangle = \delta_{ij}$, and $0\leq \omega_i \leq 1$
with the condition $\sum_i \omega_i^2 = 1$.
If $r_{\rm Schmidt}>1$ then
according to Ref.~[\onlinecite{zyczkowski98}] $|\Psi_{\rm TS}\rangle$ is inseparable and the
two blocks are entangled.

Since possible measures of entanglement for fermionic systems are the von Neumann and R\'enyi 
entropies, 
it is more efficient 
to control the truncation error~\cite{legeza02a} or the quantum information
loss $\chi$
at each renormalization step~\cite{legeza04}.
In the DMRG procedure, during the renormalization step the block $B_L$ is formed of the
subblock $B_l$ and the $l+1^{\rm th}$ site. Denoting by $s_L(l)$ the entropy of the
left subblock of length $l$ and by $s_{l+1}$ the entropy of the $l+1^{\rm th}$ site,
the change of the block entropy by forming a larger block, $B_L(l+1)$, is given as
\begin{equation}
s_L(l) + s(1)_{l+1} + I_L(l)  = s_{L}(l+1)\,,
\label{eq:information2}
\end{equation}
where 
the von Neumann entropy of a block with $l$ orbitals 
is given as 
\begin{equation}
s_L(l) = - \sum_\alpha \omega_{\alpha} \ln \omega_{\alpha},
\label{eq:s_l}
\end{equation}
where
$\omega_\alpha$ stands for
the eigenvalues of the reduced density matrix of the block $B_l$.
The so-called mutual information $I_L(l)\leq 0$ quantifies the correlation between the subsystem and the
site and it is zero if and only if the two blocks are uncorrelated.

In order to control the quantum information loss,  
$M_L$ (or $M_R$) 
is increased systematically
at each renormalization step 
until the following condition holds
\begin{equation}
s_L({l+1}) - s_L^{\rm Trunc}(l+1) < \chi\,, 
\label{eq:chi}
\end{equation}
where $\chi$ is an a priori defined error margin.  
For $s_L(l+1)$, i.e., before the truncation, $\alpha = 1\ldots M_l\times q_{l+1}$ while 
for $s_L^{\rm Trunc}(l+1)$
according to
Eq.~(\ref{eq:chi}) 
$M_L^{\rm Trunc}<M_l\times q_{l+1}$ is used.
This approach guarantees that the number of
block states are adjusted according to the entanglement between the DMRG 
blocks and the a priori defined accuracy can be reached~\cite{legeza04}.
In addition, an entropy sum rule can be
used as an alternative test of convergence~\cite{legeza04}. Therefore, we set the
minimum number of block states to $M_{\rm min}$
and $\chi$. By setting $M_{\rm min}\simeq q^3$ or $q^4$, convergence to local minima can be avoided.
In our implementation, we have $q=4$ and the basis states correspond to the 
$|0\rangle, |\downarrow\rangle,|\uparrow\rangle$ and $|\downarrow\uparrow\rangle$ states.
The maximum number of block states selected dynamically during the course of iterations
will be denoted by $M_{\rm max}$.
This approach, however, does not work for the one-site
variant of the DMRG algorithm since 
the Schmidt number of a one-site superblock configuration $M_L=M_l\times q_{l+1}$ cannot be larger 
than $M_r$. This prevents $M_l$ to
increase above $M_r$ according to Eq.~(\ref{eq:schmidt}). 
 
\subsection{Entanglement and interaction localization}

The von Neumann and R\'enyi 
entropies have been studied for quantum chemical systems as well and it was shown that
orbitals lying closer to and further away from the Fermi surface possess larger
and smaller
orbital entropy, respectively~\cite{legeza03}. The orbital entropy is related to the mixture 
of a local state and it  
is expressed  
by the eigenvalues of the reduced density matrix for a given orbital, namely
\begin{equation}
s(1)_i = - \sum_\alpha \omega_{\alpha,i} \ln \omega_{\alpha,i},
\end{equation} 
where $i=1\ldots N$ labels the orbital index while $\omega_{\alpha,i}$ stands for
$\alpha = 1\ldots q_i$
the eigenvalues of the reduced density matrix of orbital $i$.
In Fig.~\ref{fig:cu2peroxo_s1}, the single-orbital entropy is shown for the $\mu-\eta^2:\eta^2$ peroxo isomer 
of [Cu$_2$O$_2$]$^{2+}$
calculated by setting the quantum information loss $\chi=10^{-5}$.
\begin{figure}
\centerline{
\includegraphics[scale=0.68]{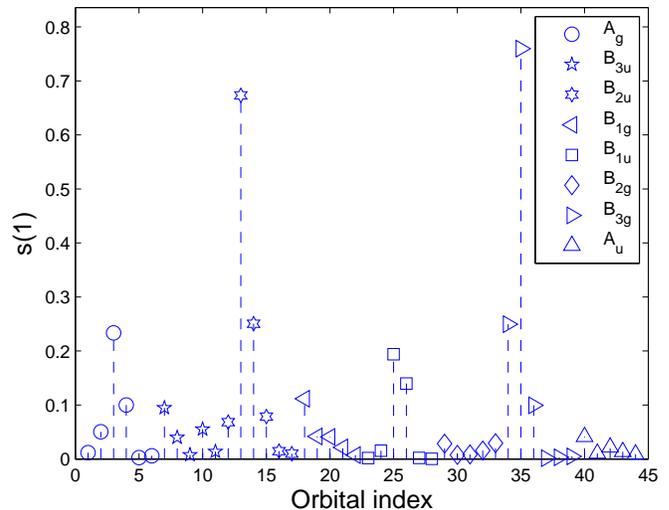}
}
\vskip .2 cm
\caption{(Color online) Single-orbital entropy function obtained 
for the $\mu-\eta^2:\eta^2$ peroxo isomer of 
[Cu$_2$O$_2$]$^{2+}$ by setting the
quantum information loss $\chi=10^{-5}$. The symbols indicate corresponding
point-group symmetries
of the energetically ordered orbitals.
\label{fig:cu2peroxo_s1}
}
\end{figure}
Orbitals with a large entropy significantly contribute to the correlation energy
whereas other slightly entangled orbitals do not.
Since the orbitals possess different single-orbital entropies, the ordering of orbitals along the
one-dimensional chain of orbitals in the DMRG algorithm has a major impact on the block entropy $s(l)$,
so that the block entropy profile can be changed based on
the ordering of the orbitals~\cite{legeza03}. 
The block entropy
also determines the required computational resources 
to reach a given accuracy~\cite{legeza03,legeza04}. 
If an optimized ordering is used, 
DMRG results can be obtained using a considerably smaller number of block states for a given
error bound.

An optimal orbital alignment by means of speed of convergence can be obtained 
by reordering the orbitals, so that
the DMRG blocks are entangled only for a few iterations. This can be achieved by
placing highly entangled orbitals at one of the ends or close to the center of
the chain.

Unfortunately, this is not true in general,  
since the independent interaction terms, like $T_{ij}$ and  
the local direct, pair-hopping, and spin-flip terms of
$V_{ijkl}$ act
as independent quantum channels; they all generate different amounts of
entanglement~\cite{legeza05a}.
Hence, localizing entanglement generated by one channel, e.g.,
the one-electron term, might lead to delocalized entanglement in another channel.
As an example, the one-electron operators are analyzed for the $\mu-\eta^2:\eta^2$ peroxo isomer of
[Cu$_2$O$_2$]$^{2+}$ shown in Fig.~\ref{fig:cu2peroxo_T}. In this case
they do not couple orbitals among different irreducible
representations of the $D_{2h}$ point group 
while the two-electron Coulomb repulsion operator does.
A proper cost function to take care of interaction and entanglement localization
can be expressed based on the two-orbital mutual information, 
\begin{equation}
I_{i,j} = s(2)_{i,j} - s(1)_i - s(1)_j,
\end{equation}
where $s(2)_{i,j}$ is the two-orbital entropy between a pair of sites
which was introduced to the QC-DMRG by Rissler {\sl et al.}~\cite{rissler06}.
If the electron--electron interactions are neglected,
the two-orbital mutual information has a similar
structure as shown in Fig.~\ref{fig:cu2peroxo_T}, where only the orbitals of the same
irreducible representations are entangled. 
The resulting single-orbital entropy, block-entropy, and the mutual information are 
shown in Fig.~\ref{fig:cu2peroxo_tij} obtained by the DMRG method after the seventh sweep. 
The block entropy oscillates
and is exactly zero when all orbitals of the same irreducible representations
belong to the left or to the right DMRG block. Therefore, the total wave function can be expressed 
as a product state of the wave functions of the subblocks of irreducible representations. 
\begin{figure}
\centerline{
\includegraphics[scale=0.45]{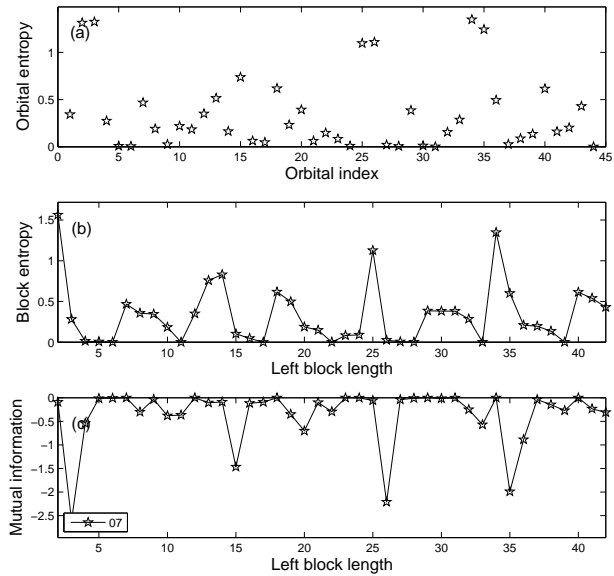}
}
\vskip -.5cm
\caption{
(Color online) (a) The single-orbital entropy, (b) block-entropy, and (c) the mutual information obtained
for the $\mu-\eta^2:\eta^2$ peroxo isomer of [Cu$_2$O$_2$]$^{2+}$ after the seventh DMRG sweep while
	neglecting the electron--electron interactions.
}
\label{fig:cu2peroxo_tij}
\end{figure}
When the two-electron integrals are also considered, orbitals among different irreducible representations
are also entangled as shown by the 
components of the two-orbital mutual information 
in Fig.~\ref{fig:cu2peroxo_I_3}
for the $\mu-\eta^2:\eta^2$ peroxo isomer of [Cu$_2$O$_2$]$^{2+}$.
\begin{figure}
\centerline{
\includegraphics[scale=0.68]{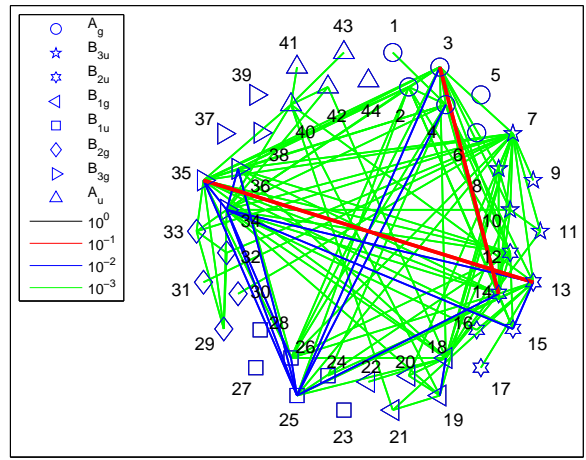}
}
\vskip -.5cm
\caption{(Color online)
Components of the two-orbital mutual information which are larger than $10^{-4}$ for the 
$\mu-\eta^2:\eta^2$ peroxo isomer of [Cu$_2$O$_2$]$^{2+}$ obtained with $\chi=10^{-5}$ and for the
energetical ordering. 
The role of symbols and numbers is the same as
in Fig.~\ref{fig:cu2peroxo_T}. 
}
\label{fig:cu2peroxo_I_3}
\end{figure}
Since the one-electron integrals are usually an order 
of magnitude larger
in chemical systems,
an optimal ordering can be found by
reordering the orbitals within the same irreducible representation and
reordering the blocks of different irreducible
representations~\cite{legeza04,legeza04b,rissler06}.
In this way,
the block entropy can be reduced significantly but the one-electron terms having
the strongest interaction remain as local
as possible. 
To draw the analogy to the Hubbard model, most
chemical systems would therefore correspond to the so-called small $U$ limit.

The reordering concept can be put in a more rigorous form by minimizing the entanglement distance,
	 which can be expressed as a cost function,
\begin{equation}
  {\hat I}_{\rm dist} = \sum_{i,j} I_{i,j} \times |i-j|^\eta\,,
\label{eq:cost}
\end{equation}
where the entanglement between pair of sites is weighted by the distance in the
chain between 
the orbitals. In Ref.~[\onlinecite{rissler06}], the effect of the parameter $\eta=-2$ has been
analyzed using simulated annealing.
Additional parameters to weight the off-diagonal elements of $I$ have been 
studied as well. In our approach, we use both $\eta=1$ and $2$. The latter choice also has the 
advantage that it can be related to the spectral algorithms of seriation problems~\cite{atkins98}.
The main aim is to sequence a set of $N$ objects, i.e., to
bijectively map the elements to the integers $1,\ldots,N$ based on   
a real valued correlation function $f(i,j)=f(j,i)$ which reflects the desire for items $i$
and $j$ to be near each other in the sequence. The two-orbital mutual
information is such a 
function which can also be seen as a weighted graph.  
In general, the problem of finding all ways to sequence the elements, so that the 
correlations are consistent, becomes NP-complete~\cite{greenberg94} in the
presence of inconsistencies. 
In such a case, there
may be no consistent solution and one needs to find the best approximation. 
If a consistent ordering is possible, the problem is {\em well posed}. Most
of the combinatorial algorithms for well-posed problems break down when the
data is inconsistent, limiting their value for many problems. 
In our approach, the minimization is performed iteratively with the constraint
that orbitals of the same irreducible 
representations are kept together, thus 
reordering of orbitals 
is allowed within
irreducible representations and 
reordering the blocks of orbitals of the 
irreducible representation is also allowed. 
As a result, a 
highly optimized ordering can be obtained as is shown 
in Fig.~\ref{fig:cu2peroxo_I_3_ord2tp}. The value for the energetical ordering  
${\hat I}_{\rm dist}=821.4$ 
is reduced to $134.1$ using $\eta=2$.  
A smaller value of 
$63$ could be reached 
by excluding the constraint discussed above. However, 
the DMRG calculations perform considerably
worse for that ordering. 

A further justification of our cost function can be discussed in terms of spectral algorithms.
Since the minimization of the cost function $g(\pi)=\sum_{ij}f(i,j)(\pi_i-\pi_j)^2$
is hard in terms of permutations $\pi$, it can be approximated by a cost function like ${\hat I}_{\rm dist}$  
of continuous variables $x_i$ that maintains its structure. 
From spectral graph theory it follows that  
the so called Fiedler vector $x=(x_1, \dots x_N)$
is the solution that minimizes $F(x)=x^\dagger L x = \sum_{ij} I_{i,j} (x_i-x_j)^2$
subject to the
following constraints that $\sum_i x_i=0$ and $\sum_i x_i^2=1$,
where the graph Laplacian is
$L=D-I$ with $D_{i,i}=\sum_j I_{i,j}$. 
The second eigenvector of the Laplacian is the Fiedler vector \cite{fiedler73,fiedler75} which 
defines a (1-dimensional) embedding of the graph on a line that tries to respect 
the highest entries of $I_{i,j}$ and the edge length of the graph. 
Ordering the entries of the Fiedler vector by non-decreasing or non-increasing way provides us
a possible ordering.
A naive application of optimization methods based on the Fiedler 
vector yielded a worse ordering than the one shown in Fig.~\ref{fig:cu2peroxo_I_3_ord2tp}.  
A more detailed analysis of energetical ordering based on the Fiedler vector
will be part of our subsequent work. 
Numerical results obtained by the optimized ordering will be further discussed
in Sec.~IV.

Note that a new electronic wave function ansatz in terms of the complete-graph
tensor network (CGTN) parametrization~\cite{marti10} might contain the same information as the
weighted graph of the cost function in Eq.\ (\ref{eq:cost}). 
This efficient parametrization might also be used to devise an optimized orbital
ordering.

\begin{figure}
\centerline{
\includegraphics[scale=0.68]{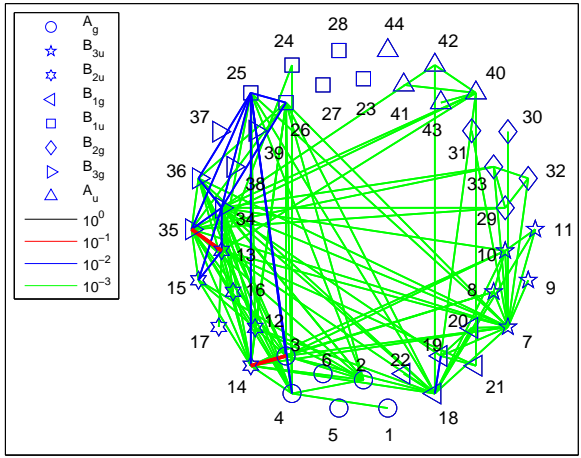}
}
\caption{(Color online)
Similar to Fig.~\ref{fig:cu2peroxo_I_3}, but for the reordered orbitals.
}  
\label{fig:cu2peroxo_I_3_ord2tp}
\end{figure}

\subsection{Efficient calculation of the single-orbital entropy}

The single-orbital entropy can be calculated for each renormalization step of
a full sweep, thus
$s(1)_i$ can be obtained for $i=1\dots N$. Therefore, the
single-orbital entropy profile for
a given ordering of molecular orbitals can be determined as a function of
sweeps~\cite{legeza03,legeza04b}. 
Once the wave function has converged by means of the entropy sum
rule~\cite{legeza04},
the single-orbital entropy profile for the given target state can be obtained.
This well
known procedure of collecting data points from subsequent renormalization steps of a full sweep
was also used by Ghosh {\sl et al.} to efficiently calculate four-point 
correlation functions~\cite{ghosh08}. 
The one-orbital entropy, on the other hand, can also be expressed in terms of 
the occupation-number representation~\cite{rissler06}.
Therefore,
the calculation of $s(1)_i$ is also possible once the required operators are 
determined for the given superblock configuration. A direct comparison of data points obtained
by the two approaches provides a reliable error estimate.

\subsection{Efficient calculation of the two-orbital entropy function}

The two-orbital entropy function can be expressed in terms of 
the occupation-number representation~\cite{rissler06}.
Unfortunately, it requires the calculation of 23 independent 
two-point correlation functions.
Since all two-point correlation functions have to be renormalized and stored
in memory or on disk due to the truncation of the superblock Hilbert space,
the efficient calculation of these operators
is crucial for feasible calculations with respect to wall time and 
computational resources.

In a standard real space DMRG procedure, the correlation functions are usually  
calculated for the symmetric superblock configuration, i.e., when the size
of the left and right blocks are equal. This configuration provides the most
accurate result for a fixed number of block states since 
the block entropy reaches its maximum value, 
so that the highest 
level of entanglement can be reached~\cite{legeza04}. 
In fact, the largest error in a 
measurable quantity is related to the largest truncation error within a 
full DMRG sweep~\cite{legeza96}. In contrast to this, 
using the DBSS method, 
the error can be kept below an a priori defined threshold,
and hence an
accurate calculation of the correlation functions is possible for the
non-symmetric superblock configurations as well. In addition, 
if two non-interacting orbitals are attached to both ends of the chain
the reduced block density matrix
at the turning points of a sweep, where
all orbitals belong to the left or the right block,
has only one non-zero eigenvalue according to the Schmidt decomposition.  
Renormalized operators required for the two-orbital entropy function 
reduce to single numbers ($M_l=M_r=1$)~\cite{legeza04,legeza04b}, consequently. 
Even without this trick, an efficient calculation of the correlation function
at the turning points is possible because the environment block
contains $q$ basis states and the system block only 
$M_{\rm min}$ states when the DBSS procedure is used. 
This is crucial in the quantum information analysis since
the construction of the mutual entropy is very costly.
In our approach, all the required operators are generated only for the 
superblock configuration at the turning points 
for which the correlation functions are calculated.
This is achieved by subsequent 
renormalization of the operators based on the a priori determined 
transformation matrices~\cite{oestlund95}.

\subsection{Configuration Interaction based Dynamically Extended Active Space (CI-DEAS) procedure}

The non-local version of the DMRG method is very sensitive to the initialization procedure.
If a poorly approximated starting configuration is used, 
the convergence can be very slow
and the DMRG can even be trapped in local minima~\cite{moritz06}. 
This can, however, be avoided by including
highly entangled orbitals from the very beginning and
expanding the active space dynamically~\cite{legeza03}.
This approach has also been extended to include protocols based on 
the Configuration Interaction (CI) procedure~\cite{legeza04b,legeza10a} and applied
to systems with sizes up to 60 orbitals.
Here, we briefly summarize the main aspects of the method as required for the
discussion in this work.

Taking a look at Fig.~\ref{fig:cu2peroxo_I_3},
some orbitals are highly entangled with several other orbitals while others are entangled with a few
orbitals only. The number of entanglement bonds emerging from the orbitals based on 
Fig.~\ref{fig:cu2peroxo_I_3} is shown in  
Fig.~\ref{fig:cu2peroxo_nl}.

\begin{figure}
\centerline{
\includegraphics[scale=0.68]{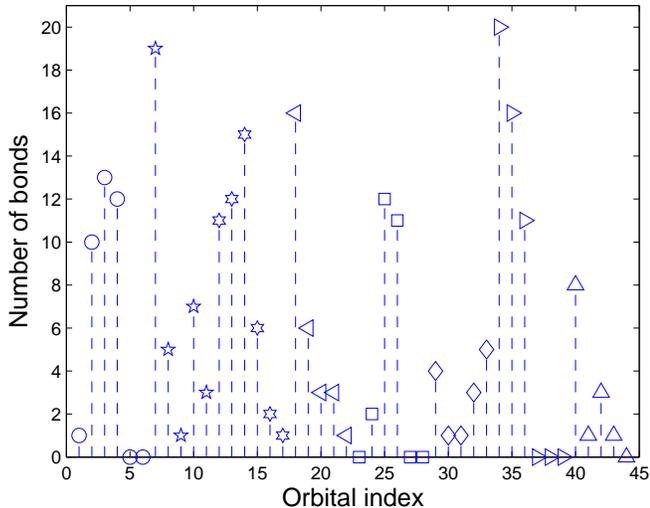}
}
\caption{(Color online)
Number of entanglement bonds emerging from the orbitals based on Fig.~\ref{fig:cu2peroxo_I_3}.  
The meaning of the symbols is the same as
in Fig.~\ref{fig:cu2peroxo_s1}.
}  
\label{fig:cu2peroxo_nl}
\end{figure}
In order to guarantee fast convergence the highly entangled orbitals --- those with several
entanglement bonds --- should be included from the very beginning of the calculations. 
If the bonds are also weighted with their strength, 
by summing $I$ column-wise,
one gets back the diagonal entries of the graph Laplacian which 
corresponds to the single-orbital entropy.
In the DEAS procedure,
   we introduce
a complete-active-space (CAS) vector 
that selects the highly entangled orbitals. The CAS vector (CASV) is formed by ordering 
orbitals with decreasing entropy
values.

Since the DMRG algorithm itself is a basis-state transformation protocol, that transforms single-particle
basis states to multi-particle basis states,
the environment block in the DEAS procedure is formed for each DMRG iteration step 
from the one-particle basis states of those orbitals which posses the
largest single-orbital entropies. 
In the first iteration step
the left block (system block) contains one orbital represented by $q$ states while
the right block (environment block) $r=N-l-2$ orbitals.
Since the exact representation
of the right block would require $q^{N-l-2}$ states which is too large for large systems,
only a subset of orbitals is included to form the active space. 
Therefore, at each DMRG iteration step of the warm up procedure, i.e., for iterations $1\ldots N-3$
the $M_r$ states are formed from those components of the CAS vector which belong to the
environment block and
possess largest entropies. The starting value of $M_r$ ($M_{\rm start}$) is set
prior of the
calculation but
during the iteration procedure $M_r$ is adjusted as $M_r = \max(M_l,M_{\rm start})$ in order to
construct at least as many environment states as the left block has in order to satisfy the
constraints set by the Schmidt decomposition.
Identifying orbitals of the right block as doubly filled, empty or active orbitals
the effective size of the environment block can be reduced significantly.
The empty orbitals can be neglected, while a partial summation over the doubly filled orbitals 
results in a constant shift in the energy. 
If DMRG auxiliary operators~\cite{xiang96}
are formed by partial summations on the left block, the effective system size of the
problem is determined by the active orbitals
only~\cite{legeza03,legeza04b,legeza10a}.
Therefore, in the warm up procedure the effective size of the system is reduced
to 5 to 7
orbitals which allows one to use larger $M_{\rm start}$ without a significant increase 
in computational time.

In order to construct even better environment states,
we also utilize CI expansion procedures.
In standard CI techniques,
the trial wave function is written
as a linear combination of determinants with expansion coefficients determined by requiring
that the energy should be minimized~\cite{jensen}. The molecular orbitals used for
building the excited Slater
determinants are taken from a CASSCF calculation and kept fix.
Therefore, in the CI method
the number of determinants included in the treatment is increased systematically
in order to achieve a better accuracy. Determinants 
which are singly, doubly, triply, quadruply, etc. excited relative
to the Hartree--Fock (HF) configuration
are indicated by the subscripts $S$,\ $D$,\ $T$,\ $Q$.
The exact wave function in a given one-particle basis, the full-CI wave
function, is then given as
\begin{equation}
\Psi_{\rm CI} = a_0 \Phi_{\rm HF} + \sum_S a_S \Phi_S + \sum_D a_D \Phi_D + \sum_T \phi_T\Phi_T + \dots
\label{eq:ci}
\end{equation}

Since the segment of the HF-orbitals belonging to the environment block is
known,
the restricted subspace of the environment block
can be formed for a given CI-level in the DEAS procedure.
Therefore, the right block contains states for a given CI-level while
the total wave function can contain higher excitations
as well due to the correlation between the two blocks.
The environment block states are constructed at each iteration step, so that the 
environment block is always optimized for the renormalized system (left) block. 
This procedure guarantees that several highly entangled orbitals are
correlated from the very beginning and both static and dynamic correlations are taken into
account. The
reduced density matrix is well defined, thus block states can be selected efficiently based
on the entropy considerations and convergence
to local minima can be avoided.
Since a significant part of the correlation energy can be obtained in this
way, usually at the end of the initialization procedure, i.e., after one-half
sweep,
chemical accuracy is often reached.
The initial CAS vector can be determined based on the chemical character of the
molecule or in a self-consistent fashion. In the latter approach, the CAS vector is set
first as CASV $= [N,N-1,\ldots,1]$ to include long range interactions between the left block
and the rightmost sites of the chain and a calculation using small values of
$M_{\rm min}$ and $M_{\rm start}$ is performed. After a full sweep,
the entropy functions are calculated and the ordering as well as the CAS vector are
determined. 
Even though the DMRG wave function has not yet converged, 
most relevant information of the system can already be extracted.
In addition, the DMRG results can be systematically improved 
using the optimized ordering and the CAS vector as a starting point for new DMRG calculations.
The CI-DEAS method allows a simple and fast calculation of all
physical quantities at the end of
the first sweep.

In the following section, the DBSS and CI-DEAS procedures are applied to chemical
systems and the results are discussed in more detail.

\section{Results}

In this section, we present results for the two isomers of [Cu$_2$O$_2$]$^{2+}$. 
We discuss our DBSS/CI-DEAS procedures by analyzing the peroxo isomer while  
for the bis($\mu$-oxo) isomer only the final results will be given.

\subsection{Electronic ground state of the two [Cu$_2$O$_2$]$^{2+}$ cores}

Here, we summarize the main steps of our entropy-based optimization procedures.
First, we perform a short DMRG calculation using a limited set of
block states in order to obtain the most important characteristics of the entropy
functions and to determine the low-lying energy spectrum.
In order to use DMRG in an automated manner as a black-box method,
we do not use any specific ordering
or CAS vector related to the chemical characteristics of the problem.

In order to determine the multiplicity of the converged target state,
we also calculate the expectation value of the $S^2$ operator given as
\begin{equation}
S^2 = \sum_{ij} S^-_i S^+_j + \sum_{ij} S^z_i S^z_j + \sum_i S^z_i\,.
\end{equation}
The expectation value is equal to $S(S+1)$ in Hartree atomic units, i.e., 
zero for a singlet state and two for a triplet state.

We performed a calculation with fixed $M=64$ states, using the
energetical ordering,
and the CAS vector was simply set to CASV$\equiv [N, N-1, \ldots, 1]$. This latter choice 
guarantees that during the initialization procedure  
long range interaction between the system block (left-block)
and the rightmost sites of the chain is taken into account.
In this calculation, we have restricted the CI-DEAS procedure to include
determinants with at most triple excitations from the Hartree--Fock state.
The convergence of the ground state energy
in the $A_g$ irreducible representation of the point group $D_{2h}$ is shown in 
Fig.~\ref{fig:cu2peroxo_energy}(a) as the square symbols.  
\begin{figure}
\centerline{
\includegraphics[scale=0.45]{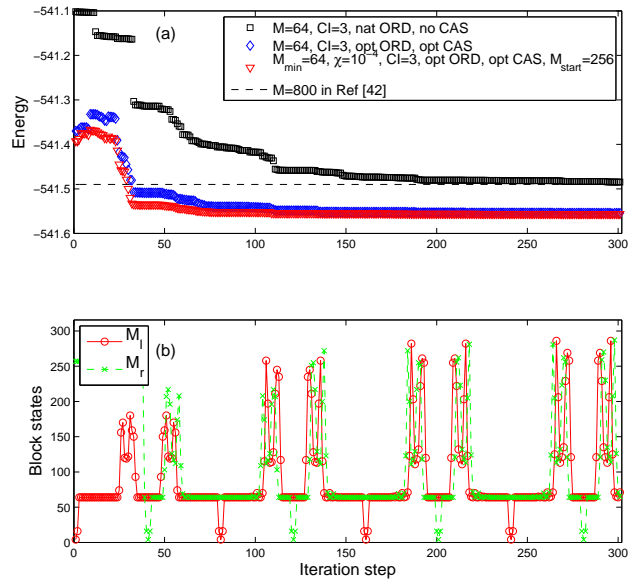}
}
\caption{(Color online)
(a) Convergence of the ground state energy in Hartree for various parameters for the
$\mu-\eta^2:\eta^2$ peroxo isomer. The
dashed line gives the ground state energy obtained in Ref.~[\onlinecite{marti08}] using 
$M=800$ block states.
(b) Number of block states as a function of iteration steps
for $\chi=10^{-4}$ corresponding to Fig.~\ref{fig:cu2peroxo_ord2tp_entropy}. 
\label{fig:cu2peroxo_energy}
}
\end{figure}
Although the convergence is rather slow,   
the results obtained in our previous work with $M=800$ block states~[\onlinecite{marti08}]
-- shown by the dashed line -- can be reached after eight sweeps. 
The convergence to the singlet state has been confirmed by the expectation value of
the total-spin operator $\langle S^2\rangle = 10^{-3}$.

The block entropy profile 
converges very slowly as shown in Fig.~\ref{fig:cu2peroxo_ord1_ci1}(b). In fact, the
entropy profile corresponding to the CI-DEAS procedure represented by the red circle symbol
is almost zero for most of the superblock configurations which 
indicates the lack of entanglement between the DMRG blocks.
As a consequence, the DMRG algorithm selected inappropriate states for the
description of the environment blocks.

A detailed analysis of the entropy
functions provides even better convergence properties. 
\begin{figure}
\includegraphics[scale=0.45]{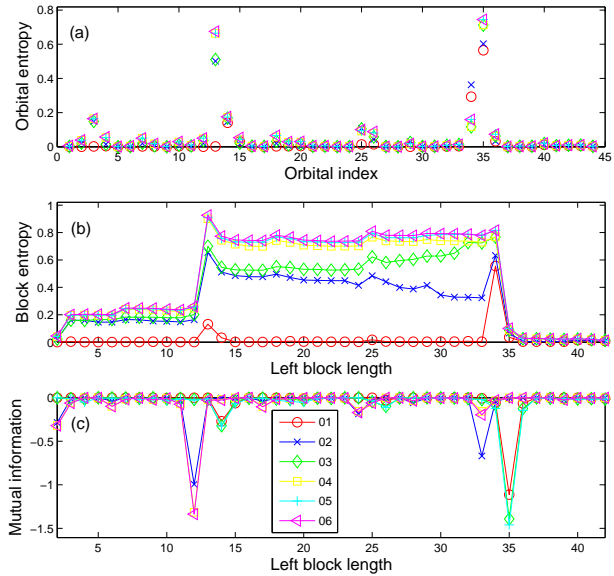}
\vskip .2 cm
\caption{(Color online)(a) Single-orbital entropy, (b) block entropy and (c) mutual information profile 
for the $\mu-\eta^2:\eta^2$ peroxo isomer of 
[Cu$_2$O$_2$]$^{2+}$ 
using energetical ordering, without optimized CAS vector for the singlet ground
state of $A_g$ 
symmetry with fixed $M=64$ block states.
The convergence of the ground state energy in Hartree is shown in Fig.~\ref{fig:cu2peroxo_energy}(a)
by the square symbol. 
\label{fig:cu2peroxo_ord1_ci1}
}  
\end{figure}
The obtained single-orbital and block entropies   
as a function of DMRG sweeps are shown in Fig.~\ref{fig:cu2peroxo_ord1_ci1}.
By comparing Figs.~\ref{fig:cu2peroxo_ord1_ci1}(a) and \ref{fig:cu2peroxo_s1},
   we see
that orbitals with large entropies can already be identified and a reasonable CAS vector
can thus be constructed. 
In Fig.~\ref{fig:cu2peroxo_ord1_ci1}(b), one also recognizes that once the third orbital
is pushed in the left block, i.e., for $l\ge 3$ the block entropy increases significantly. 
This is because the third orbital is highly entangled with the 14$^{\rm th}$ and 25$^{\rm th}$
orbitals as can be seen in Fig.~\ref{fig:cu2peroxo_I_3}. 
The largest values of the block entropy is reached for $13<l<35$ when the highly entangled orbitals 
$13$ and $35$
belong to the two different blocks. 
In order to localize entanglement,
ordering optimization based on the two-orbital mutual information 
can also be carried out according to the procedure described in Sec.~II and   
a block entropy profile shown 
in Fig.~\ref{fig:cu2peroxo_ord2tp_entropy}(b) is obtained.
\begin{figure}
\centerline{
\includegraphics[scale=0.45]{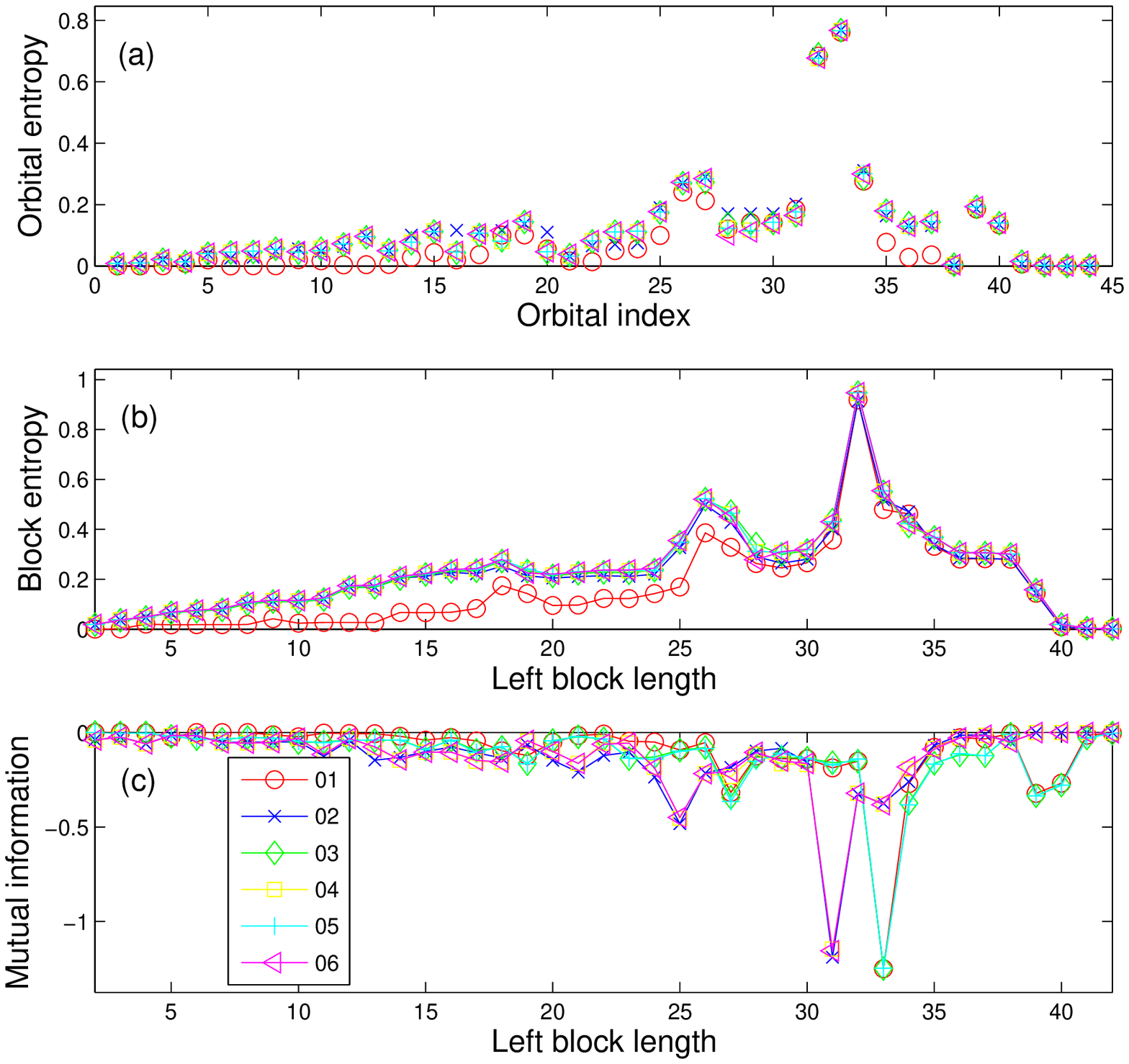}
}
\caption{(Color online)(a) Single-orbital entropy, (b) block-entropy, and (c) mutual information obtained for 
the $\mu-\eta^2:\eta^2$ peroxo isomer of [Cu$_2$O$_2$]$^{2+}$ as a function of
	DMRG sweeps with optimized 
ordering, CAS vector, and by setting the
quantum information loss $\chi=10^{-4}$. The CI-DEAS initialization procedure corresponds
to symbols with red circle. The optimized ordering and CAS vector is given in Table~\ref{tab:cu2peroxo}.
\label{fig:cu2peroxo_ord2tp_entropy}
}
\end{figure}
By comparing Figs.~\ref{fig:cu2peroxo_ord1_ci1}(b) and \ref{fig:cu2peroxo_ord2tp_entropy}(b) 
it is clear that in the latter case the block entropy is highly localized, i.e., it
takes large values only for a few superblock configurations. 
Repeating the calculation with 
fixed $M=64$ 
block states again but using an optimized ordering and CAS vector yields a
ground state energy estimate of
$E_{\rm peroxo}=-541.55533$ Hartree. 
The fast convergence is shown in Fig.~\ref{fig:cu2peroxo_energy} 
by the diamond symbol. In fact, the ground state energy after the first half sweep, i.e.,
at the end of the CI-DEAS procedure is already far below the one given by Ref.~[\onlinecite{marti08}]
and the block entropy profile has the same structure as the one obtained after
eight sweeps.   
This can be attributed to
the CI-DEAS procedure for selecting the appropriate environment block states 
at each iteration step
by taking care of the renormalized system block and the inclusion of the highly
entangled orbitals from the beginning.

There is no need to use additional methods
to guarantee convergence like white noise~\cite{chan02} or perturbative
correction~\cite{kura09} which is an interesting observation in view of the
results of Kurashige and Yanai.
The optimized ordering and CAS vectors used in the calculation are
summarized in Table~\ref{tab:cu2peroxo}. 
\begin{table}
\begin{tabular}{ccccccccccccccccc}
\hline
ORD = [&  44 & 41 & 42 & 43 & 40 & 31 & 30 & 33 & 32 & 29 & 11 & 10 &  9 &  8 &  7 \\
      & 20 & 21 & 19 
      & 18 & 22 &  1 &  2 &  5 &  6 &  4 &  3 & 14 & 12 & 17 & 16 \\
      & 15 & 13 & 35 & 34 & 36 & 38 
      & 37 & 39 & 25 & 26 & 24 & 27 & 28 & 23 ] \\
\\
CASV = [& 35 & 13 & 14 & 34 &  3 & 25 & 26 & 18 &  4 & 36 &  7 & 15 & 12 & 10 & 2\\
       &19 & 40 & 20 &  8 & 33 & 29 & 42 & 21 & 32 & 24 & 16 & 11 & 43 &1 & 41 \\
       & 17 & 44 & 31 & 30 &  9 & 22 &  6 & 39 & 38 &  5 & 27 & 23 &37 & 28]\\

\hline
\end{tabular}
\caption{Optimized ordering and CAS vector for the $\mu-\eta^2:\eta^2$ peroxo isomer of [Cu$_2$O$_2$]$^{2+}$.}
\label{tab:cu2peroxo}
\end{table}
Carrying out the same procedure for the bis($\mu$-oxo) isomer of
[Cu$_2$O$_2$]$^{2+}$,
the convergence of the ground state energy with fixed $M=64$ states 
without optimized ordering and CAS vector is shown
in Fig.~\ref{fig:cu2bisoxo_energy}(a) by the square symbol. 
We also observe for the bis($\mu$-oxo) isomer, that our DMRG energies reproduce
the results given in Ref.~[\onlinecite{marti08}] with a fraction of the computational resources.
\begin{figure}
\centerline{
\includegraphics[scale=0.45]{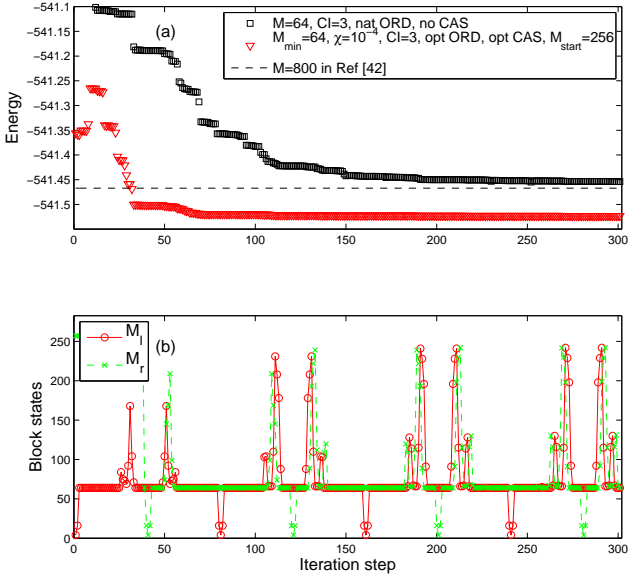}
}
\caption{(Color online)
(a) Convergence of the ground state energy in Hartree for various parameters
for the bis($\mu$-oxo) isomer.
The dashed line gives the ground state energy obtained in Ref.~[\onlinecite{marti08}] using 
$M=800$ block states.
(b) Number of block states as a function of iteration steps
for $\chi=10^{-4}$.
\label{fig:cu2bisoxo_energy}
}
\end{figure}
\begin{figure}
\centerline{
\includegraphics[scale=0.68]{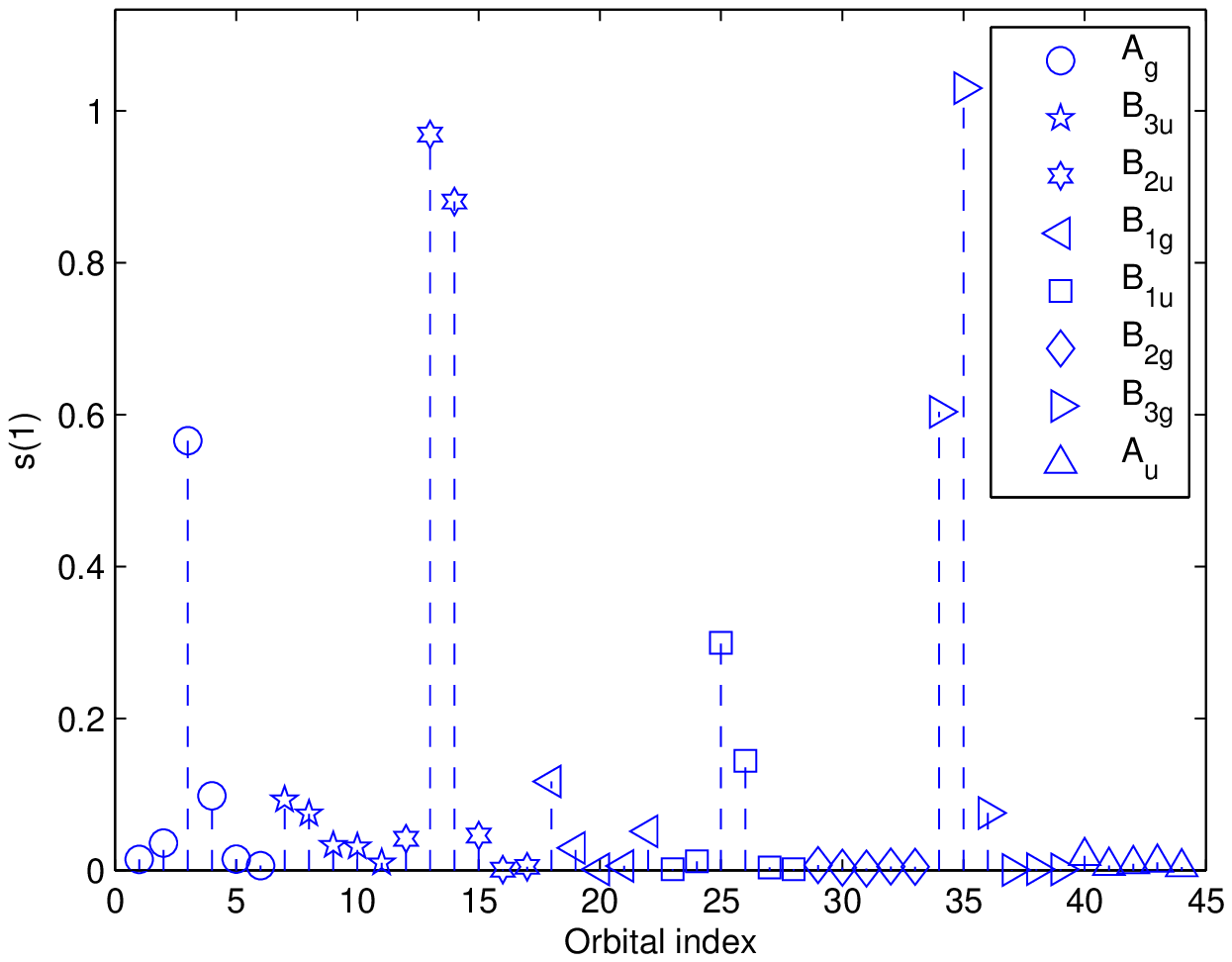}
}
\vskip .2 cm
\caption{(Color online)Similar to Fig.~\ref{fig:cu2peroxo_s1} but for the bis($\mu$-oxo) isomer of [Cu$_2$O$_2$]$^{2+}$.
\label{fig:cu2bisoxo_s1}
}
\end{figure}
It is clear that the convergence of the energy is very slow to the value given in
Ref.~[\onlinecite{marti08}], however, after 
the entropy-based optimization procedures, a much faster convergence to a much lower
value can be reached.
As a result, the optimized CAS vector based on the single-orbital entropy shown in Fig.~\ref{fig:cu2bisoxo_s1}
is given in Table~\ref{tab:cu2bisoxo}. 
\begin{table}
\begin{tabular}{ccccccccccccccccc}
\hline
ORD = [&44 & 41 & 43 & 40 & 42 & 31 & 30 & 33 & 32 & 29 & 11 & 10 &  9 &  8 &  7 \\
       &20 & 21 & 19 & 18 & 22 &  1 &  2 &  5 &  6 &  4 &  3 & 12 & 17 & 16 & 15 \\
       &14 & 13 & 35 & 34 & 36 & 38 & 37 & 39 & 25 & 26 & 24 & 27 & 28 & 23 ]\\
\\
CASV = [&35 & 13 & 14 & 34 & 3  & 25 & 26 & 18 &  7 & 4  & 8  & 36 & 22 & 15 & 12 \\
       &2  & 9  & 10 & 19 & 40 & 5  & 42 & 1  & 11 & 29 & 24 & 43 & 44 & 33 & 41 \\
       &32 & 30 & 6  & 21 & 17 & 31 & 27 & 20 & 28 & 38 & 23 & 29 & 16 & 37 ]\\
\hline
\end{tabular}
\caption{Optimized ordering and CAS vector for the bis($\mu$-oxo) isomer of [Cu$_2$O$_2$]$^{2+}$.}
\label{tab:cu2bisoxo}
\end{table}
The highly localized two-dimensional entanglement matrix for the optimized
ordering is understood from the comparison of 
Figs.~\ref{fig:cu2bisoxo_I_3} and \ref{fig:cu2bisoxo_I_3_ord4}. 
The optimized ordering vector utilized in the calculation is given in Table~\ref{tab:cu2bisoxo}.
The entanglement distance ${\hat I}_{\rm dist}=1043$ for the
energetical ordering is reduced to ${\hat I}_{\rm dist}=102.5$ for the optimized ordering.
\begin{figure}
\centerline{
\includegraphics[scale=0.68]{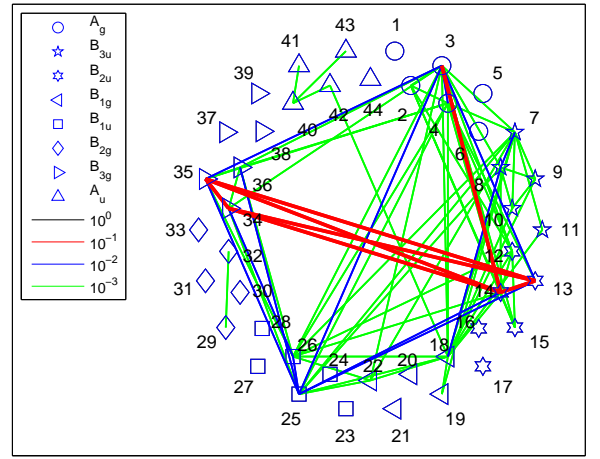}
}
\caption{(Color online)
Similar to Fig.~\ref{fig:cu2peroxo_I_3}, but for the bis($\mu$-oxo) isomer of [Cu$_2$O$_2$]$^{2+}$.
\label{fig:cu2bisoxo_I_3}
}  
\end{figure}
The resulting entropy profiles as a function of sweeps are shown in Fig.~\ref{fig:cu2bisoxo_ord4_entropy}.
Again the block entropy takes large values only for a few iteration steps within a full sweep. 
\begin{figure}
\centerline{
\includegraphics[scale=0.68]{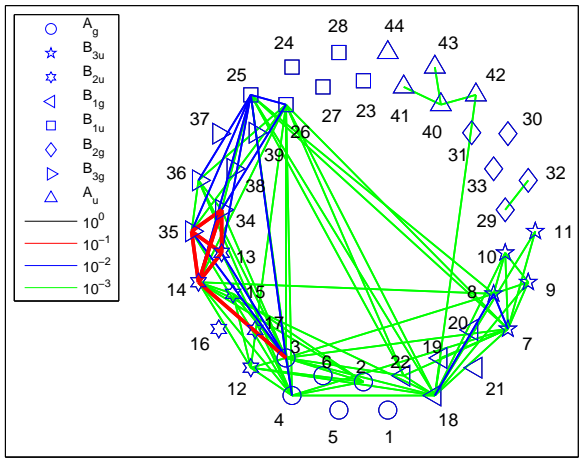}
}
\caption{(Color online)
Similar to Fig.~\ref{fig:cu2bisoxo_I_3}, but for the optimized ordering.
\label{fig:cu2bisoxo_I_3_ord4}
}  
\end{figure}
\begin{figure}
\centerline{
\includegraphics[scale=0.45]{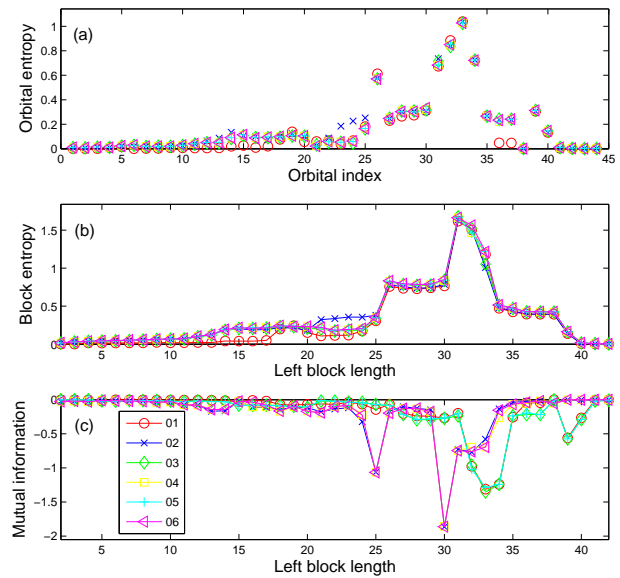}
}
\caption{(a) Single-orbital, (b) block-entropy, and (c) mutual information obtained for 
bis($\mu$-oxo) isomer of [Cu$_2$O$_2$]$^{2+}$ as a function of DMRG sweeps with optimized 
ordering and CAS vector and for setting the
quantum information loss to $\chi=10^{-4}$. The CI-DEAS initialization procedure corresponds
to symbols with red circle. The optimized ordering and CAS vector is given in the text.
}
\label{fig:cu2bisoxo_ord4_entropy}
\end{figure}

Since a very important observable related to transition metal clusters is the
energy difference between the two isomers, the ground state energies should
be calculated for the same error margin. This cannot be guaranteed with a fixed
number of block states but it can be achieved by the DBSS procedure (or an
automated Richardson-type error protocol~\cite{marti2010a}). 
We have thus calculated the ground state energies of the two isomers using the
same parameter set, namely for $M_{\rm min}=64$, $\chi=10^{-4}$, $M_{\rm start}=256$.
The convergence of the ground state energy
is plotted in Fig.~\ref{fig:cu2peroxo_energy}(a) by the triangle symbol, while
the number of block states selected dynamically 
in Fig.~\ref{fig:cu2peroxo_energy}(b).  
Since the 
block entropy is highly localized for the optimized ordering 
for most of the superblock configurations the number of
block states were determined by $M_{\rm min}$ as 
can be seen in Fig.~\ref{fig:cu2peroxo_energy}(b) and  
Fig.~\ref{fig:cu2bisoxo_energy}(b).
We found 
$E_{\rm peroxo}=-541.55628$ Hartree and $E_{\rm bisoxo}=-541.51416$ Hartree 
yielding an energy differences of $0.04212$ Hartree, i.e., $110$~kJ/mol. 
For both isomers we have obtained                                                                       
$\langle S^2\rangle =10^{-4}$ for the ground state                                                      
as expected from the given error margin.  

The convergence
of the ground state energy for the bis($\mu$-oxo) isomer is shown in Fig.~\ref{fig:cu2bisoxo_energy}(a)
by the triangle symbol. The fast convergence is again evident. 
The effective size of the environment block
during the first $N$ iteration steps,
i.e., for the CI-DEAS procedure, was found to be less than five orbitals.
Therefore, using a larger number 
of block states for the environment block ($M_{\rm start}$) increases the computational time
insignificantly.  
This result indicates that the application of DBSS and CI-DEAS procedures
in the entropy-based optimization
guarantees that the DMRG algorithm can be used in a black-box fashion, and that
chemical accuracy can be reached using very limited computational resources.
In fact, the reduced effective system size in the CI-DEAS procedure allows
one to obtain the most relevant characteristics of the entropy functions within a few
minutes.

To ensure that during the CI-DEAS procedure an even better represented
environment blocks are constructed, we have repeated the calculations using
$M_{\rm min}=256$  and $M_{\rm start}=1024$. After eight sweeps, we obtained the following
$E_{\rm peroxo}=-541.57900$ Hartree and $E_{\rm bisoxo}=-541.53599$ Hartree with 
a gap of $113$ kJ/mol.
In addition, we have performed more accurate calculations 
for $\chi=10^{-5}$, $M_{\rm min}=256$, and $M_{\rm start}=512$. 
The maximum number of block states selected dynamically was around 1000 for the
peroxo isomer, while  
for the bis($\mu$-oxo) isomer we found that slightly more states were required
to reach the same accuracy, so that $M_{\rm max}$ was in the range of 1200.
In both calculations, 
we obtained $\langle S^2\rangle=10^{-5}$ as expected for a singlet state. 
For the given error margin, we found $E_{\rm peroxo}=-541.58050$ Hartree, 
$E_{\rm bisoxo}=-541.53565$ Hartree, 
thus the difference is $0.04485$ Hartree
that is $118$ kJ/mol. 

In Table~\ref{tab:sum}, all relevant DMRG calculations for the bis($\mu$-oxo) and $\mu-\eta^2:\eta^2$ peroxo
isomers of the binuclear copper cluster are listed.
\begin{table}
{\small
\begin{tabular}{lrrr}
\hline\hline
Ref.,method & $E_{\rm bisoxo}$ &  $E_{\rm peroxo}$ & $\Delta{E}$\\ 
\hline
\multicolumn{4}{c}{Reference energies} \\
\hline
\cite{cram06},CASSCF(16,14)    & -541.50307 & -541.50345 & 1\\
\cite{cram06},CASPT2(16,14)    & -542.06208 & -542.06435 & 6\\
\cite{cram06},bs-B3LYP         & -544.19419 & -544.27844 & 221 \\
\cite{malmqvist2008},RASPT2(24,28)& &                       & 120 \\
\hline
\multicolumn{4}{c}{Previously published DMRG energies} \\
\hline
\cite{marti08},DMRG(26,44)[800]     & -541.46779 & -541.49731 &  78 \\
\cite{marti2010b},DMRG(26,44)[128]  & -541.47308 & -541.51470 & 109 \\
\cite{kura09},DMRG(32,62)[2400]     & -541.96839 & -542.02514 & 149 \\
\cite{yanai10},DMRG(28,32)[2048]-SCF    & -541.76659 & -541.80719 & 107 \\
\cite{yanai10},DMRG(28,32)[2048]-SCF/CT &            &              & 113 \\
\hline
\multicolumn{4}{c}{DMRG energies from this work} \\
\hline
DMRG(26,44)[64/256/$10^{-4}$]         & -541.51416 & -541.55628 & 111 \\
DMRG$^\ast$(26,44)[256/512/$10^{-4}$] & -541.53499 & -541.57896 & 115 \\
DMRG$^\ast$(26,44)[256/1024/$10^{-4}$]& -541.53599 & -541.57900 & 113 \\
DMRG(26,44)[256/512/$10^{-5}$]        & -541.53565 & -541.58050 & 118 \\
\hline\hline
\end{tabular}
}
\caption{
The relative energies for the bis($\mu$-oxo) and $\mu-\eta^2:\eta^2$ peroxo [Cu$_2$O$_2$]$^{2+}$ isomers
obtained in this work and from previously published studies are listed in kJ/mol.
All total energies are given in Hartree.
The CAS is denoted in parentheses as '(electrons,orbitals)'
while the information on DMRG block states $M$ is given in brackets.
Note that total energies for different CAS and different types of orbitals cannot be
directly compared.
All results from this work employ entropy-based optimization applying the DBSS
and CI-DEAS procedures. 
The ``$^\ast$'' denotes keeping slightly more DMRG environment
states during the initialization than necessary for a given quantum
information loss $\chi$.
The square brackets [M$_{\rm min}$/M$_{\rm start}$/$\chi$] state that the DMRG
calculation starts with M$_{\rm start}$
block states, and the minimal number of block states is set to M$_{\rm min}$, respectively.
\label{tab:sum}
}
\end{table}
Total electronic energies of the two isomers reported by Malmqvist~\emph{et al.}~\cite{malmqvist2008} 
and Kurashige and Yanai~\cite{kura09} were calculated in a different active space.
The latter authors then carried out new DMRG calculations \cite{yanai10}, where they reduced
their active space, but applied the canonical-transformation (CT) approach
\cite{yanai06} and optimized the orbitals.
A comparison of absolute energies is therefore not very meaningful as the total
energy depends on the size of the CAS and the type of orbitals chosen.
Instead we shall focus on the relative energy which is central to the chemistry
of such systems.
Our entropy-based DMRG calculations agree quantitatively
with the RASPT2(24,28) and DMRG-SCF/CT calculations, even though no procedure --- such as
canonical transformation --- 
has been employed to account for dynamical correlation in our studies. 
However, this is not surprising because Yanai {\it et al.} found \cite{yanai10}
that the effect of CT is only about 4 kJ/mol in the given one-particle basis set.
In the first study~\cite{marti08} on this problem we already found 
the relative DMRG energies
converge faster than the absolute energies of each isomer. 
This finding has been confirmed by Yanai~\emph{et al.}~\cite{yanai10}.

It is evident from Table~\ref{tab:sum} that already our first DMRG estimate~\cite{marti08} for this relative
energy was much closer than the CASSCF result to the correct splitting, 
which we may expect between $110$ and $160$ kJ/mol 
based on the RASPT2 calculation and the Kurashige--Yanai DMRG result for a much
larger active space. Our improved result for the splitting~\cite{marti2010b}, 
which we obtained for the original active space but with a reduced number 
of block states $M_l=128$ is with $109$ kJ/mol already very close to the
DMRG-SCF/CT and RASPT2 results
of $113$ and $120$ kJ/mol, respectively.
Even the total energies turned out to be improved and in the case of the peroxo
copper cluster below the small-CAS CASSCF result as it should be.
In view of the fact that a discrepancy in relative energy of about $5$ to $10$ kJ/mol 
is acceptable for chemical accuracy, we emphasize that both results, 
$113$ and $118$ kJ/mol, are close and that hence already the small DMRG calculation 
is a feasible means to yield such relative energies.

\subsection{Entropic analysis of the two isomers}

Besides the relative energies of the two isomers, 
one might deduce more information from the entropy functions 
related to the chemical properties of the binuclear copper cores.
By comparing Figs.~\ref{fig:cu2peroxo_s1} and
\ref{fig:cu2bisoxo_s1}, it is clear that the importance of the orbitals is
different for the two isomers as shown by the different distribution of the
single-orbital entropy. 
In general, almost all orbitals possess some 10-20\% larger entropy for the
bis($\mu$-oxo) isomer but some of the orbitals have 2.5 -- 3 times larger entropy 
than those of the peroxo isomer.
For example, orbitals 3, 14, 34 produce much larger contributions
to the total entanglement in case of the bis($\mu$-oxo) isomer. 
The total quantum correlation encoded in  
the ground state,  $I_{\rm Tot}=\sum_{i=1,\ldots N} s(1)_i$, is $3.49$ and
$5.39$ for the 
$\mu-\eta^2:\eta^2$ peroxo and bis($\mu$-oxo) isomers, respectively. Thus,
the bis($\mu$-oxo) isomer is more entangled which is also reflected 
by the larger   
maximum
value of the block entropy for the bis($\mu$-oxo) isomer as 
can be seen by comparing 
Figs.~\ref{fig:cu2peroxo_ord2tp_entropy} and \ref{fig:cu2bisoxo_ord4_entropy}.  

\begin{figure*}
\centerline{
\includegraphics[scale=0.5]{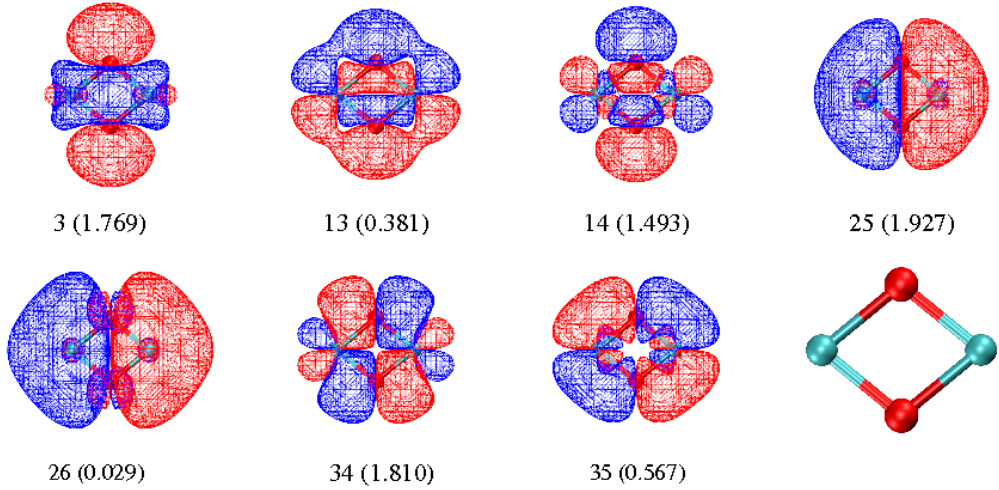}
}
\caption{(Color online)
Molecular orbital pictures of the highly entangled orbitals for the bis($\mu$-oxo)
	isomer. 
The orbitals were
selected with respect to their one-site entropy as shown in
Fig.~\ref{fig:cu2bisoxo_s1}.
The number below each orbital corresponds to the orbital index and the
occupation number is written in the parentheses.
\label{fig:orbs_bisoxo}
}  
\end{figure*}
\begin{figure*}
\centerline{
\includegraphics[scale=0.5]{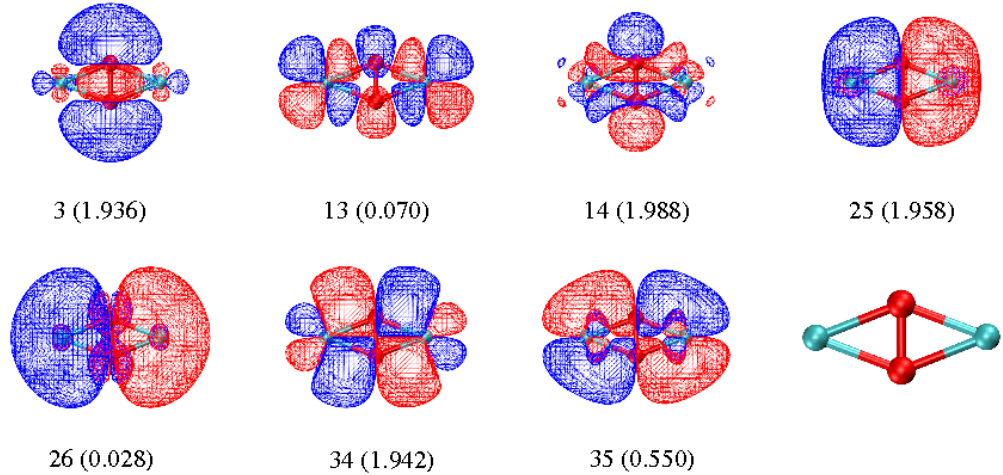}
}
\caption{(Color online)
Molecular orbital pictures of the highly entangled orbitals for the peroxo
isomer. 
The orbitals were
selected with respect to their one-site entropy as shown in
Fig.~\ref{fig:cu2peroxo_s1}.
The number below each orbital corresponds to the orbital index and the
occupation number is written in the parentheses.
\label{fig:orbs_peroxo}
}  
\end{figure*}

The highly entangled molecular orbitals for the $\mu-\eta^2:\eta^2$ peroxo and bis($\mu$-oxo) isomers are shown in
Figs.~\ref{fig:orbs_bisoxo} and \ref{fig:orbs_peroxo}.
The molecular orbital analysis shows that the highly entangled orbitals have an
occupation number that strongly deviates from either being doubly occupied or
empty. The entropy-based optimization scheme accurately determines those orbitals 
and thus allows one to perform efficient DMRG calculations with the smallest
possible active space for a desired accuracy in the DMRG energy.
The selected orbitals are close to the Fermi surface and would be included in a
standard CASSCF calculation if one could employ these large active spaces as in
DMRG calculations.

\section{Summary and Perspective}

\subsection{Relation to Tensor Network States}

Our procedure has been demonstrated on a one-dimensional spatial topology related to the
DMRG method, while the obtained two-dimensional entanglement matrix could be
employed more efficiently in methods based on 
tensor network state (TNS) approaches.
For example,  
the two-orbital mutual information can provide 
an optimal value for the coordination number of each orbital
in case of the tree tensor network state (TTNS) approach~\cite{murg10}
and   
an optimized network topology can be determined.
Since in this latter method the distance between highly entangled orbitals
can be reduced significantly compared to the DMRG topology 
entropy-based optimizations are expected to boost its 
convergence properties to a great extent. As an example, 
we present calculations performed with the TTNS method on the Be atom
studied recently in Ref.~[\onlinecite{murg10}]. 
The components of the two-orbital mutual information larger than $10^{-4}$ are
shown in Fig.~\ref{fig:be48_I_3}(a). The convergence of the ground state energy 
with a fixed bond dimension $D=2$ and coordination number $z=3$ for three  
different topologies shown in Fig.~\ref{fig:ttns} is plotted 
in Fig.~\ref{fig:be48_I_3}(b).  
\begin{figure}
\centerline{
\includegraphics[scale=0.55]{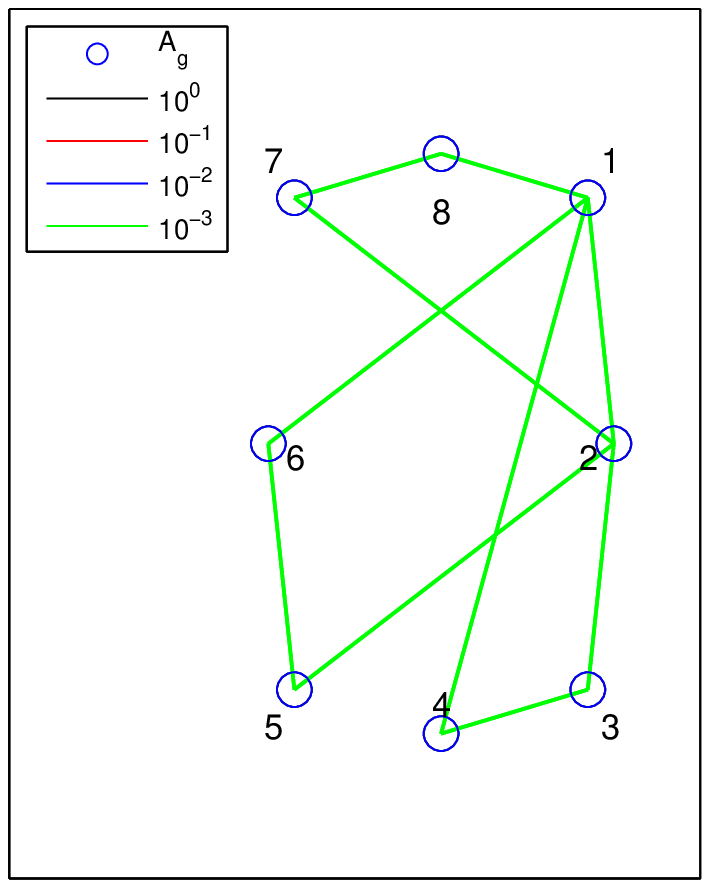}
\includegraphics[scale=0.55]{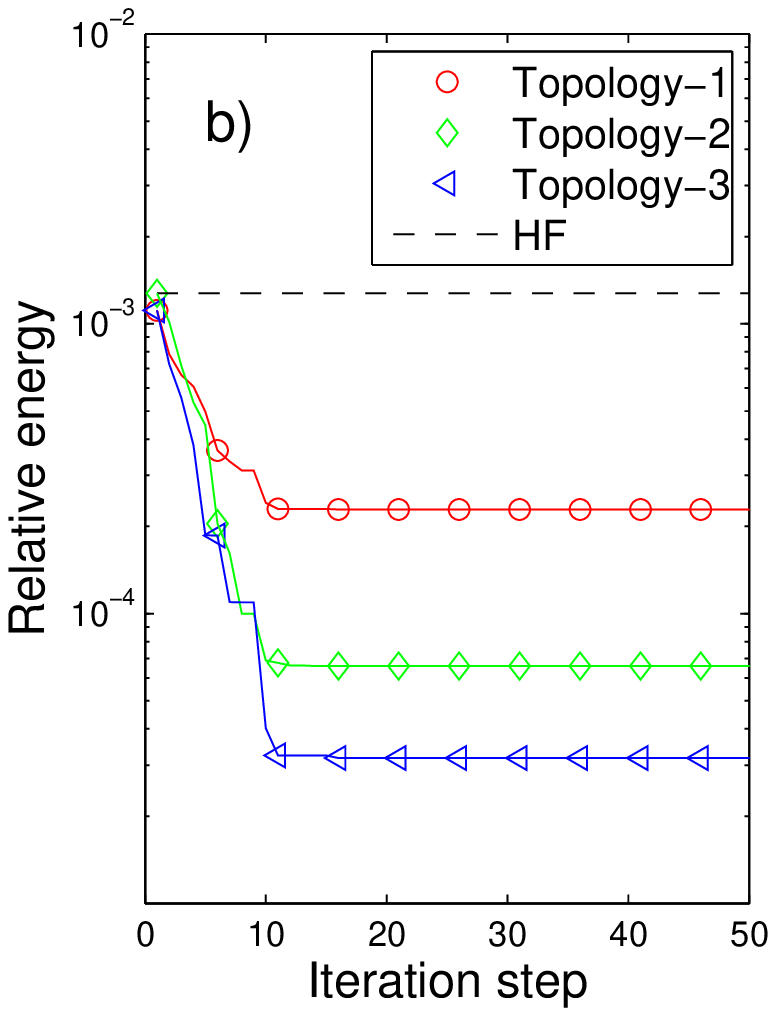}
}
\caption{(Color online) (a) Similar to Fig.~\ref{fig:cu2peroxo_I_3}, but for the Be atom. 
(b) Convergence of the ground state energy for the Be atom obtained with the TTNS method with 
a fixed bond dimension $D=2$ and coordination number $z=3$ for three different network topologies
shown in Fig.~\ref{fig:ttns}~\cite{murg2010priv}.
\label{fig:be48_I_3}
}  
\end{figure}
\begin{figure}
\centerline{
\includegraphics[scale=0.4]{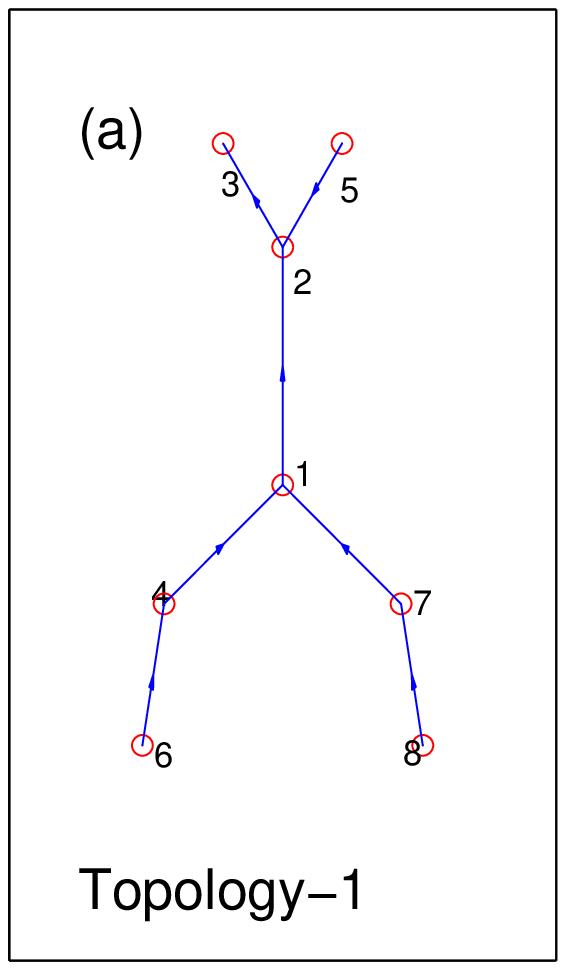}
\includegraphics[scale=0.4]{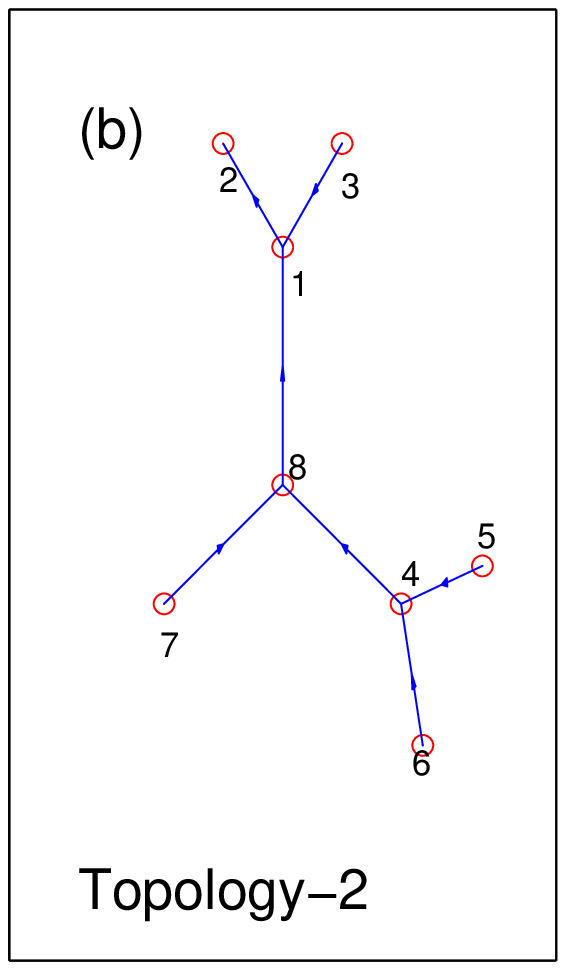}
\includegraphics[scale=0.4]{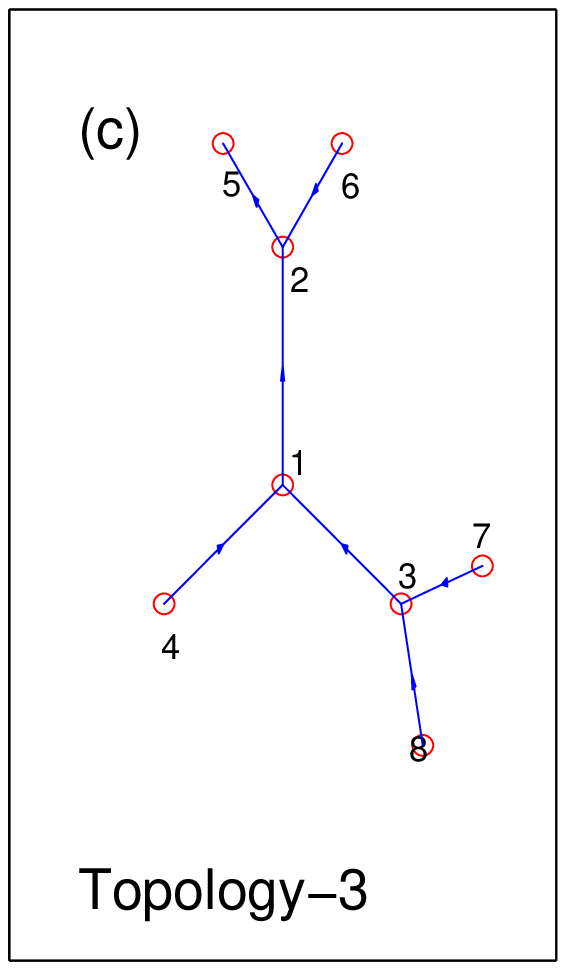}
}
\caption{(Color online)
Three different network topologies were used for the TTNS method. The topology shown in (c)  
is optimized based on the two-orbital mutual information.
\label{fig:ttns}
}
\end{figure}
It is found that for this set-up the relative
error in the ground state energy dropped by an order of magnitude if an optimized topology
based on the two-orbital mutual information is used. For the one-dimensional 
DMRG topology and for the energetical ordering, we found that ${\hat I}_{\rm dist}=0.7251$
whereas it is reduced to $0.236$ for the optimized topology of the TTNS method.
For the topologies shown in Fig.~\ref{fig:ttns} (a) and (b) we found 
a considerably larger values.
We want to emphasize that for the one-dimensional DMRG topology the lowest value of
${\hat I}_{\rm dist}$ is $0.30$ by including all permutations of the orbitals.
For the optimized ordering based on the 
Fiedler vector, 
${\hat I}_{\rm dist}$ is $0.319$ for which, in fact,   
we obtained the lowest ground state energy ($E_{rel}=1.69\times 10^{-5}$) when fixed  
$M=8$ DMRG block states were used.
A detailed study based on spectral analysis and graph theory and extension of
our approach to larger systems is under progress and will be part of a subsequent work.  

The optimization of the spatial arrangement of orbitals and 
network topologies has a significant 
influence on the convergence properties of MPS and TNS algorithms. 
However, the total entanglement encoded in the wave function, $I_{\rm Tot}$, cannot be changed.
The optimization of the one-particle basis can yield entanglement reduction
in the system and the value of $I_{\rm Tot}$ can be manipulated, consequently. 
In fact, as discussed in Ref.~[\onlinecite{rissler06}] by constructing an optimal basis it 
might be that ordering is either obvious or irrelevant.
In the past few years, 
various procedures have been developed for  
orbital optimizations~\cite{white02,yanai06,yanai10,zgid08,zgid08a,ghosh08,luo10,murg10}.
We will therefore continue to examine mutual information with respect to orbital
optimization in our subsequent work. 

\subsection{Conclusion}
In this paper, we have studied
a transition metal cluster from a quantum information
theory perspective using the DMRG method. 
By calculating various entropy functions and the two-orbital mutual information
we have proposed recipes
to perform DMRG calculations in a black-box fashion
on complex chemical compounds. 

Optimizations based on the two-orbital mutual information can be related to
graph theory and spectral analysis of seriation problems.
Our cost function is interpreted in terms of
the Fiedler vector of the graph Laplacian.
Our results confirm the importance of taking entanglement among molecular
orbitals into account and the usefulness of graph theory for carrying out efficient calculations.
Chemical characteristics of the two isomers of [Cu$_2$O$_2$]$^{2+}$
have also been analyzed and interpreted in terms
of the entropy functions. 

The present work confirms our previous findings
that even small-$M$ DMRG calculations provide a qualitatively correct 
description of transition metal clusters as demonstrated in
Ref.~\cite{marti08,marti2010b}.

Our results indicate that highly entangled orbitals form subgroups. 
Therefore, a coarse-graining approach might be possible which 
could be efficiently implemented by the multiscale-entanglement-renormalization
ansatz (MERA)~\cite{vidal06}.

In future work, we shall explore the promising DBSS/CI-DEAS method in DMRG
calculations on a set of test molecules featuring states whose relative energy
is difficult to calculate. Also, additional options for improvement like extrapolation
schemes~\cite{marti2010a} shall be investigated.

\acknowledgments{
This work was supported in part by
Hungarian Research Fund (OTKA) through Grant Nos.~K68340 and
K73455.
\"O. L. acknowledges support from
the Alexander von Humboldt foundation. 
K. M. and M. R. gratefully acknowledge financial support
through a TH-Grant (TH-26 07-3) from ETH Zurich.
We thank A. Mai, R. M. Noack, and Cs. Nemes for very
valuable discussions and V. Murg for performing calculations with the 
TTNS method on the Be atom.
We would especially like to thank 
the Erwin-Schr\"odinger-Institut in
Vienna for its hospitality during the Quantum Computation and Quantum
Spin Systems workshop in 2009,
the ETH 
Zurich and the Pauli Center for Theoretical Studies at ETH Zurich 
for its hospitality during the CECAM 
workshop of ``Tensor network methods for quantum chemistry'' in 2010, 
where many fruitful discussion took place.
}

\end{document}